# Preferential functionalization on zigzag graphene nanoribbons: First-principles calculations

Hoonkyung Lee

Department of Physics, University of California at Berkeley, Berkeley, California 94720, USA

Materials Sciences Division, Lawrence Berkeley National Laboratory, Berkeley, California 94720, USA

Corresponding author: hkiee3@civet.berkeley.edu.

ABSTRACT

We investigate the functionalization of functional groups to graphene nanoribbons with zigzag and armchair edges using first principles calculations. We find that the formation energy for the configuration of the functional groups functionalized to the zigzag edge is ~0.2 eV per functional group lower than that to the armchair edge. The formation energy difference arises from a structural deformation on the armchair edge by the functionalization whereas there is no structural deformation on the zigzag edge. Selective functionalization on the zigzag edge takes place at a condition of the temperature and the pressure of ~25 $^{o}$C and $10^{-5}$ atm. Our findings show that the selective functionalization can offer the opportunity for an approach to the separation of zigzag graphene nanoribbons with their solubility change.

# 1. Introduction

Since graphene which is a single layer of graphite was made in 2004[1], graphene has been of great interest because of unusual physical properties such as being the thinnest metal, massless Dirac fermion, and having half-integer and fractional quantum Hall effects, and high mobility[1-4]. Recently, it has been observed that the bandgap of bilayer graphene with Bernal stacking (i.e., AB stacking) can be widely tunable up to ~0.25 eV by applying a gate bias[5], which well agrees with a first-principles calculation's result[6]. These studies show that nanoelectronics and nanophotonics based on graphene is feasible. However, the gate bias is needed for opening bandgaps and single or more than triple graphene layers are metallic and their bandgaps are not tunable with a bias, which is an obstacle for applications.

On the other hand, graphene nanoribbons (GNRs) which are composed of one dimensional honeycomb $sp^2$ carbon with zigzag or armchair edges have also attracted a great deal of attention because of their potential applications to device materials.[7-11] It has been shown that zigzag-edged graphene nanoribbons (ZGNRs) and armchair-edged graphene nanoribbons (AGNRs) can be used for nanoelectric and nanophotonic devices because of their bandgaps of ~0.5–3.0 eV. The bandgaps for ZGNRs and AGNRs arise from the antiferromagnetic interaction between magnetic moments on the opposite edges and quantum confinement effects, respectively.[12-16] Another notable feature of GNRs is that ZGNRs can have half-metallicity through time reversal symmetry breaking by an external electric field.[17] These studies show that ZGNRs can be utilized for electronics as well as spintronics based on graphene. Recently, GNRs have been synthesized experimentally from unzipping carbon nanotubes (CNTs) by chemical, plasma, and electric breakdown methods.[18-20] Moreover, progress on making GNRs with regular edges has been made through unzipping CNTs by a sonification method[21] and field effect transistors (FET) with GNRs have been implemented[8]. These experiments show the feasibility of nanoelectrics and spintronics based on graphene. However, the practical applications of ZGNRs for a nanoscale spintronics face the separation of ZGNRs similarly as in the case of the separation of semiconducting carbon nanotubes (CNTs).

Since ZGNRs and AGNRs have different edge shapes, it is expected that the selective functionalization of functional groups on either edge of ZGNRs and AGNRs may be possible. In this paper, we find that the formation energy for the configuration of the functional groups functionalized to the zigzag edge is ~0.2 eV per functional group lower than that to the armchair edge. The energy difference arises from a structural deformation on the armchair edge by the functionalization unlike in the case of the zigzag edge. Selective functionalization on the zigzag edge takes place at a condition of the temperature and the pressure of ~25 $^o$C and $10^{-5}$ atm. The selective functionalization can offer the opportunity for an approach to the separation of zigzag graphene nanoribbons with their solubility change.

## 2. Computational details

All of our calculations were performed using first principles density functional calculations with the Vienna Ab-initio Simulation Package (VASP). The exchange correlation energy functional of generalized gradient approximation (GGA) for Perdew and Wang[22] was employed, and the kinetic energy cutoff was taken to be 400 eV. For the electron-ion interaction, the projector augmented wave (PAW) method was used. The optimized atomic positions were obtained by relaxation until the Hellmann-Feynman force on each atom was less than 0.01 eV/Å. Supercell[23] calculations were employed throughout where the atoms on adjacent nanoribbons were separated by over 10 Å to eliminate unphysical interactions between image structures on the periodic calculations.

## 3. Results and discussion

We perform the functionalization of functional groups on the edges of ZGNRs and AGNRs in order to investigate whether the functional groups prefer to be functionalized to either edge of ZGNRs and AGNRs. We choose five different kinds of functional groups: hydroxyl (−OH), amine (−NH$_2$), phenyl (−C$_6$H$_5$), carbonyl (−COOH), and cyano (−CN) groups as well as a ZGNR with a width of 13.5 Å and an AGNR with a width of 17.8 Å. In our model system, one hydrogen atom of the ZGNR and AGNR is replaced with one functional group as described in Figs. 1(a) and 1(b), respectively. The distance between the functional group and the H atom in the ZGNR (AGNR) for hydroxyl, amine, phenyl, carbonyl, and benzyl groups is 2.6 (1.9), 2.5 (2.2), 2.6 (2.1), 2.5 (2.0), and 3.1 (2.2) Å, respectively. For AGNRs, the functionalization results in somewhat of a distortion of the edge carbon regime close to the functional groups whereas there is not any structural deformation on ZGNRs. The structural deformation on AGNRs by the functionalization is ascribed to the steric hindrance between the functional group and the nearest hydrogen atom of the ZGNR.

To investigate the selective functionalization of the functional groups between ZGNRs and AGNRs, the formation energy per functional group for the configuration of the functional groups functionalized to ZGNRs and AGNRs is calculated as follows:

$$F = E[M\text{-}GNR] + E[H_2] + \mu_{H_2} - E[GNR] - E[HM] - \mu_{HM} \quad (1)$$

where E[X] indicates the total energy of the systems for X, M-GNR and HM stand for the functionalized GNR and a hydrogen passivated functional group, respectively, and $\mu_{H_2}$ and $\mu_{HM}$ are the chemical potentials of H$_2$ gas and a MH phase, respectively. We find that the formation energy for the functionalized ZGNRs is lower than that for the functionalized AGNRs by ~0.2 eV per functional group regardless of the kinds of the functional groups. We attribute the lower formation energy for the configuration of the functionalized ZGNRs to having no structural deformation on ZGNRs by the functionalization while there is a structural deformation on AGNRs. Since the chemical potential ($\mu_{HM}$) for the MH phase can be neglected because the chemical potential is very small if a long molecule containing the functional groups is chosen, the formation energy for the functionalized ZGNR and AGNR can be written by $F_i^{ZGNR} = \mu_{H_2} + a_i^{ZGNR}$ and $F_i^{AGNR} = \mu_{H_2} + a_i^{AGNR}$, respectively, where $i$ indicates the kinds of functional groups. The calculated coefficients ($a_i^{ZGNR}$ and $a_i^{AGNR}$) of the formation energy for the functionalized ZGNRs and AGNRs are presented in Table I. When $F_i^{ZGNR} < 0$ and $F_i^{AGNR} > 0$ (i.e., $-a_i^{AGNR} < \mu_{H_2} < -a_i^{ZGNR}$), the selective functionalization to ZGNRs takes place. For instance, at a condition of the temperature and the pressure of 25 °C and $10^{-5}$ atm or 230 °C and 1 atm ($\mu_{H_2} \approx -0.6$ eV from ideal gas), the functionalized ZGNRs are stable whereas the functionalized AGNRs are unstable except for in the case of the functionalization of the hydroxyl and cyano groups. At 25 °C and 10 atm ($\mu_{H_2} = -0.24$ eV), the cyano-functionalized ZGNRs are stable whereas the cyano-functionalized AGNRs are unstable. At 90 °C and $10^{-5}$ atm ($\mu_{H_2} = -0.73$ eV), the hydroxyl-functionalized ZGNRs are stable whereas the hydroxyl-functionalized AGNRs are unstable. These results show that the selective functionalization of functional groups to ZGNRs is feasible.

To investigate the interaction between the functional groups attached on the edge of ZGNRs, we have done supercell calculations on the total energy for the configuration of two functional groups attached to the edge of a ZGNR as the distance between the functional groups changes from 4 to 13 Å. We find that the configuration for the functional groups attached to the edge of the ZGNR where they attach with a distance of more than ~12 Å is the lowest in energy except for in the case of the hydroxyl group as shown in Fig. 2(a). This implies that the interaction between the functional groups is repulsive to each other, and the minimum distance with negligible interaction between them is ~12 Å. In contrast, the interaction between the hydroxyl groups is attractive to each other because of the attractive electrostatic interaction between the hydroxyl groups by the electron accumulation on the O atom and the electron depletion on the H atom as shown in Fig. 2(b). Therefore, the estimated coverage of the functional groups on the edge of ZGNRs for hydroxyl, amine, carbonyl, phenyl, and cyano groups is 100, 50, 20, 20, and 20%, respectively. We have also confirmed that the interaction between the functional groups attached to the opposite edges of ZGNRs is negligible.

We study the electronic structures for the functionalized ZGNRs to examine whether the functionalization influences their electronic and magnetic structures. The antiferromagnetic ground state for hydrogen- (or pristine), hydroxyl-, amine-, carbonyl-, phenyl-, and cyano-functionalized ZGNRs with a width of 13.5 Å is lower than the ferromagnetic ground state by 13.0, 8.6, 3.8, 3.0, 7.4, and 5.1 meV per edge carbon atom, respectively. Figure 3 shows the bandstructure of the hydrogen- and hydroxyl-functionalized ZGNRs for comparison. The calculated bandgaps of the hydrogen-, hydroxyl-, amine-, carbonyl-, phenyl-, and cyano-functionalized ZGNRs with the coverage of 100, 100, 50, 20, 20, and 20% are 0.41, 0.31, 0.38, 0.40, 0.39, and 0.39 eV, respectively. These results imply that the functionalization does not affect the electronic and magnetic properties.

We have shown the possibility of the selective functionalization of functional groups on ZGNRs for its application for the separation of ZGNRs with their solubility change. For instance, hydroxyl-functionalized graphene nanoribbons are polar while unfunctionalized ones are nonpolar, such that the functionalized and unfunctionalized graphene nanoribbons are soluble and nonsoluble in a polar solvent, respectively, or vice versa in a nonpolar solvent. Recent experiments[24,25] have shown that functional groups prefer to be functionalized on the edges of graphene, and the functionalization causes their solubility change, which supports our approach. The coverage of the edge of ZGNRs by the functionalization can reach approximately as much as 20–100%, which may be enough to change the solubility of the functionalized ZGNRs. Furthermore, according to a recent theoretical paper,[26] the functionalization of functional groups (e.g., hydroxyl group) on the edges of ZGNRs enhance the half-metallicity. Therefore, functionalized ZGNRs can be directly applicable as device materials. These experimental and theoretical results indicate that the separation of ZGNRs by the selective functionalization we propose here can be feasible by the solubility difference between functionalized ZGNRs and unfunctionalized AGNRs.

## 4. Conclusion

In conclusion, we have shown the feasibility of selective functionalization of functional groups on the edge of zigzag graphene nanoribbons using first principles density-functional calculations. This selective functionalization would be utilized for an approach to the separation of zigzag graphene nanorribons with their solubility change. Our results also may indicate that zigzag graphene nanoribbons made from unzipping carbon nanotubes by a chemical method is favorable.

**ACKNOWLEDGMENT.** H. L. was supported by NSF Grant No. DMR07-05941. Numerical simulations were supported by the Director, Office of Science, Office of Basic Energy Sciences, Materials Sciences and Engineering Division, U. S. Department of Energy under Contract No. DE-AC02-05CH11231. Computational resources were provided by NERSC and TeraGrid.

**Table I**. Coefficients ($a_i^{ZGNR}$ and $a_i^{AGNR}$) of the formation energy for the functionalized ZGNR and AGNR as the functional groups from $F_i^{ZGNR} = \mu_{H_2} + a_i^{ZGNR}$ and $F_i^{AGNR} = \mu_{H_2} + a_i^{AGNR}$, respectively. $\Delta F = F_i^{ZGNR} - F_i^{AGNR}$ ($= a_i^{ZGNR} - a_i^{AGNR}$).

| Functional groups ($i$) | $a_i^{ZGNR}$ (eV) | $a_i^{AGNR}$ (eV) | $\Delta F$ (eV) |
|---|---|---|---|
| -OH | 0.72 | 0.89 | −0.17 |
| -NH$_2$ | 0.53 | 0.76 | −0.23 |
| -COOH | 0.51 | 0.73 | −0.22 |
| -C$_6$H$_5$ | 0.57 | 0.79 | −0.22 |
| -CN | −0.04 | 0.26 | −0.30 |

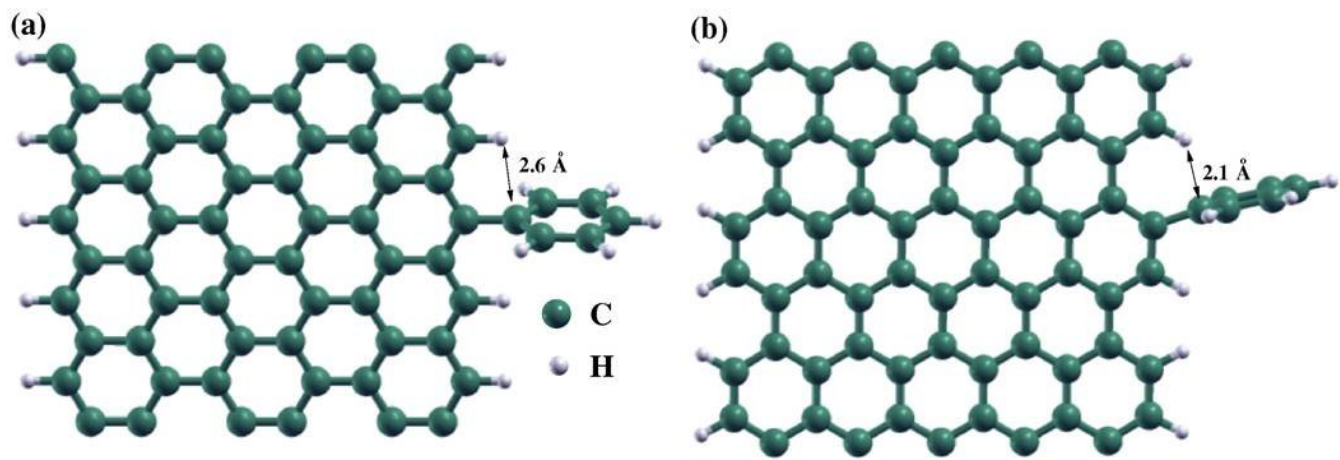

**Figure 1.** (a) and (b) The optimized atomic geometries for a phenyl group attached to the edge of a ZGNR and an AGNR, respectively. The distance between the C atom of the phenyl group and the H atom in a ZGNR and an AGNR is 2.6 and 2.1 Å, respectively.

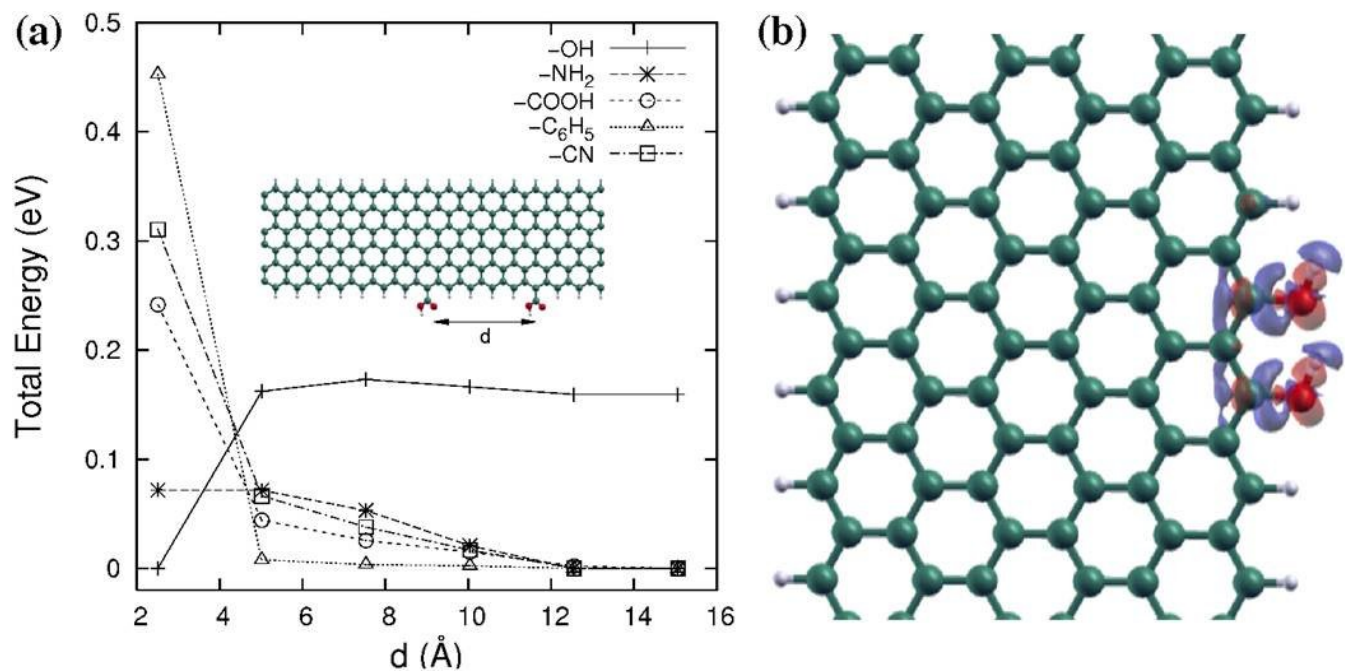

**Figure 2.** (a) The total energy of the optimized structures for two functional groups attached to the edge of the ZGNR as a function of different initial positions where the energy of the lowest energy configuration is set to zero. The inset shows the optimized atomic geometry for two carbonyl groups attached to the edge of a ZGNR. (b) The charge density difference between the O atom and the H atom or the ZGNR with the isosurface value of 0.01 $e/(au)^3$ when two hydroxyl groups are attached to the edge of the ZGNR. Red and blue colors indicate electron accumulation and depletion, respectively.

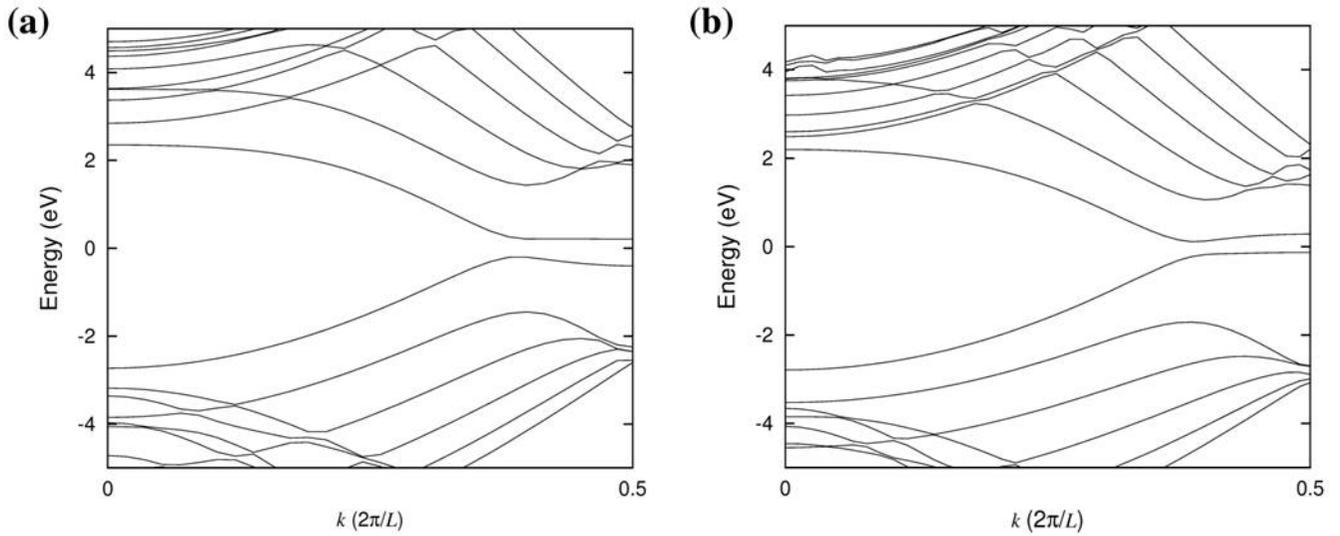

**Figure 3.** (a) and (b) The bandstructures for a ZGNR with a width of 13.5 Å and a hydroxyl-functionalized ZGNR (100% edge coverage), respectively. *L* is the in-plane lattice constant.